\documentclass{article}
\usepackage[utf8]{inputenc}
\usepackage{authblk}
\usepackage{url}
\usepackage{hyperref}
\usepackage{graphicx}
\usepackage{booktabs}
\usepackage{xcolor}

\newcommand{\mytitle}{Assessing disinformation through the dynamics of supply and demand in the news ecosystem}

\title{\mytitle}

\author[a]{Pietro Gravino*}
\author[a]{Giulio Prevedello} 
\author[a]{Martina Galletti}
\author[a,b,c]{Vittorio Loreto}

\affil[a]{Sony Computer Science Laboratories, 75005 Paris, France}
\affil[b]{Sapienza University of Rome, Physics Department, 00185, Rome, Italy}
\affil[c]{Complexity Science Hub Vienna, A-1080 Vienna, Austria}
\affil[*]{Corresponding author: pietro.gravino@sony.com}

\begin{document}
\maketitle

\begin{abstract} 
Social dialogue, the foundation of our democracies, is currently threatened by disinformation and partisanship, with their disrupting role on individual and collective awareness and detrimental effects on decision-making processes. Despite a great deal of attention to the news sphere itself, little is known about the subtle interplay between the offer and the demand for information. Still, a broader perspective on the news ecosystem, including both the producers and the consumers of information, is needed to build new tools to assess the health of the infosphere. Here, we combine in the same framework news supply, as mirrored by a fairly complete Italian news database - partially annotated for fake news, and news demand, as captured through the Google Trends data for Italy. Our investigation focuses on the temporal and semantic interplay of news, fake news, and searches in several domains, including the virus SARS-CoV-2 pandemic. Two main results emerge. First, disinformation is extremely reactive to people’s interests and tends to thrive, especially when there is a mismatch between what people are interested in and what news outlets provide. Second, a suitably defined index can assess the level of disinformation only based on the available volumes of news and searches. Although our results mainly concern the Coronavirus subject, we provide hints that the same findings can have more general applications. We contend these results can be a powerful asset in informing campaigns against disinformation and providing news outlets and institutions with potentially relevant strategies.
\end{abstract}



\section*{Introduction}
The Covid-19 crisis evidenced once more that disinformation stands as one of the major plagues of the Information Age. In the last decades, many national and international institutions started to implement a vast plethora of strategies to tackle this issue~\cite{funke2018guide} and mitigate its effects. Still, the mechanisms underlying the role and phenomenology of disinformation are largely unclear.

Only in recent times the complex ecosystem of information massively attracted the interest of the scientific community. Disinformation went under investigation, from its very definition~\cite{fallis2015disinformation} to its psychological mechanisms~\cite{bakir2018fake}, and its spreading dynamics~\cite{cinelli2020covid}. Detection and forecast of disinformation were also among the relevant topics explored by the scientific community~\cite{shu2017fake}. These studies raised questions about how to identify statistical markers in the news content~\cite{conroy2015automatic} or about the diffusion mechanisms~\cite{vicario2019polarization}. 

A meaningful part of the research effort focused on the impact of disinformation on diverse fields of human activities, such as consumers' behaviour~\cite{visentin2019fake}, political elections~\cite{allcott2017social}, sustainability~\cite{treen2020online} or health~\cite{kata2010postmodern}. During the Covid-19 pandemic, particularly, the effect of disinformation on social behaviours became so compelling that the term ``Infodemic'' made a comeback from the SARS epidemic of 2003~\cite{rothkopf2003buzz}, to describe the spreading of false or incorrect information about the virus SARS-CoV-2. The consequences were disastrous~\cite{islam2020covid} and led to dangerous behaviours that further aggravated the epidemic crisis. 

While disinformation is always under the spotlight, the complex ecosystem of information, which is the substrate for disinformation, attracted much less interest.  It is important to stress that the infosphere relies on the subtle interplay of two types of actors: news producers on the one hand and news consumers on the other. In this structure, the supply and the demand of information stand in a market-like relationship. The study of their interplay is essential to unveil the mechanisms of information dynamics. It also provides a broader view in which disinformation can be contextualised and analysed.

The news supply can be identified with the overall news production, mainly consisting of officially recognised newspapers. The general news production had been primarily studied in linguistics~\cite{dafouzlinguistics}, while analyses of news content~\cite{korobchinsky2017peculiarities} and coverage~\cite{schmidt2013media,sznitman2015cannabis} were often focusing on particular countries or topics. Other works investigated the impacts of news and its consumption on, for example, reading behaviour~\cite{tewksbury2003americans}, finance~\cite{engle1993measuring}, and political opinions~\cite{harteveld2018blaming}.

News demand, instead, is more difficult to pinpoint. In the literature, surveys and lab studies are usual procedures of investigation~\cite{trussler2014consumer, tewksbury2003americans, iyengar2004consumer}, but, unlike general news production, they cannot scale up to the population level. Thus, different solutions have to be adopted. The tracking of reading behaviours, for example, had been used to study the demands and interests of readers~\cite{boczkowski2011choice}. However, such a methodology is biased by the very existence of news since the interest for topics not covered by news cannot be recorded.

An independent way to track people interests that gained popularity in the scientific community is the Google Trends service\footnote{\url{https://trends.google.com/}}~\cite{jun2018ten}. It provides an index proportional to the number of searches made with the Google Search engine, enabling the quantitative comparison of searched queries. In the last decade, Google Trends has been mainly used as a marker, and a predictor, of people's behaviours in different contexts, like finance~\cite{da2015sum,preis2013quantifying}, epidemiology~\cite{lampos2015advances,dugas2013influenza} or socio-economic indicators~\cite{choi2012predicting,borup2020search}. Interestingly, its intrinsic value as a proxy for people's interest was perhaps overlooked. In the framework of news, the Google Trends index has been mainly adopted for forecasting~\cite{weeks2010symbiosis}, without delving into the comprehension of the dynamics of the news ecosystem. 

Here we comprehend, in a unique framework, the supply and demand for information and analyse their dynamical interplay with the final goal of understanding the main mechanisms of the information ecosystem dynamics and extracting hints about the determinants of disinformation. To this end, we focused on the general production of news in Italy, from early December 2019 to the end of August 2020, as the reference for the news supply. For the same period, the Google Trend index served as a proxy for the general public's information demand.

We adopted Vector Auto-Regression (VAR) models to study the interplay between news demand and supply, evidencing different causal relationships for distinct subjects. We presented an improved modelling scheme that allows for a quantitative description of the dependencies in the time series evolution for information demand and supply. 

The new framework also permitted to study and compare the disinformation dynamics within the general information system, highlighting behavioural differences in reactivity and modelling efficacy. Furthermore, we observed that the semantic misalignment between information supply and demand is higher than the misalignment between disinformation supply and demand. 

These discrepancies could be exploited to aggregate a disinformation risk indicator that is independent of fake news annotations. We contend this index could provide a reliable and independent assessment tool for the news supply's health status. 


\section*{Results}

\subsection*{Dynamics of news supply and demand}

Information systems feature two main drivers: news supply and news demand. As a reference for the news supply, we looked at the whole Italian production of information, termed General News, from early December 2019 to the end of August 2020. For the same period, the Google Trend index served as a proxy for the news demand from the Italian general public, thus termed Searches (refer to Materials and Methods for more details).

\begin{figure*}[ht]
\centering
\includegraphics[width=1.\textwidth]{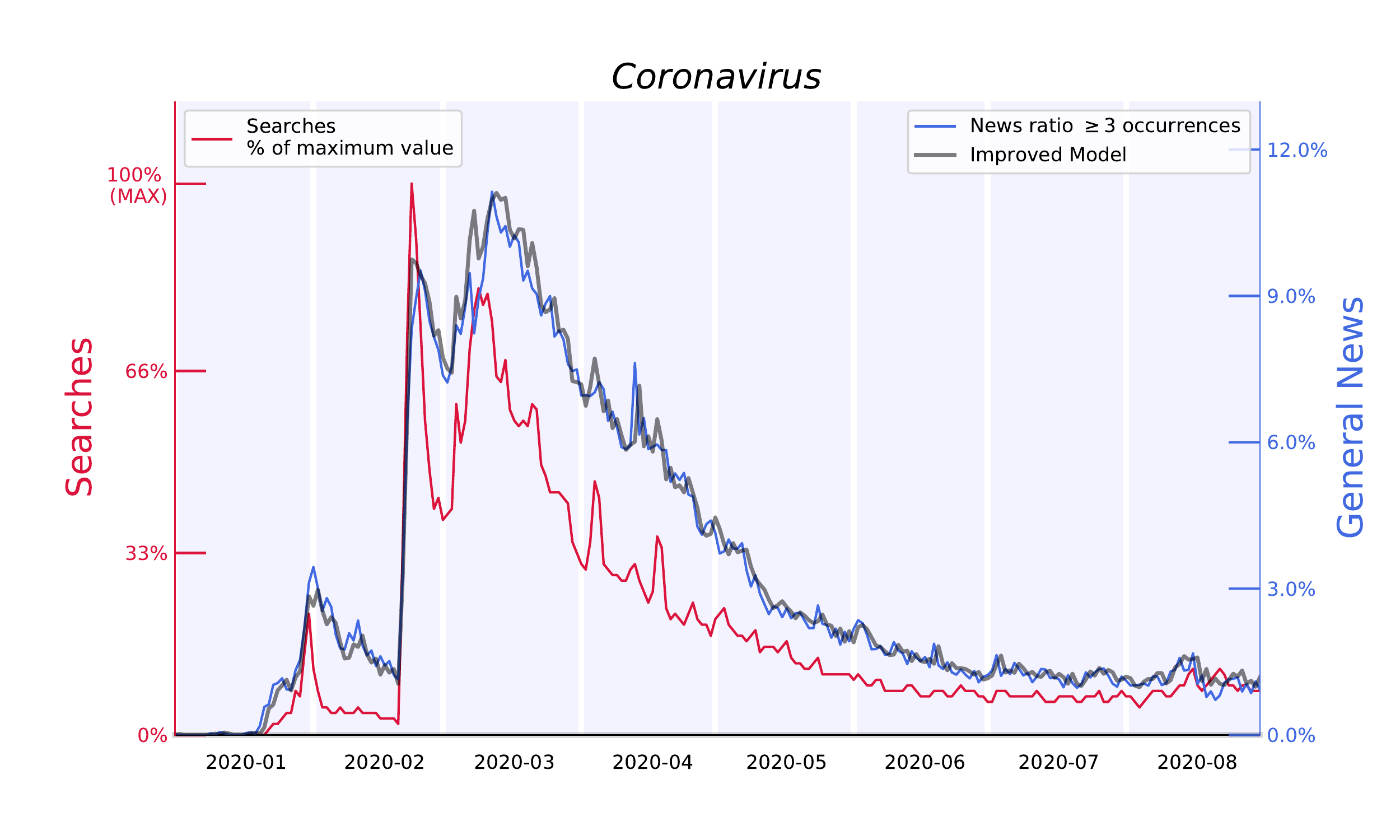}
\caption{Temporal behaviour of the fractions of Searches (red, left $y$-axis) and General News (blue, right $y$-axis) for the keyword \emph{coronavirus} in Italy from early December 2019 to the end of August 2020. Searches are reported as a percentage of the maximum observed in the monitored period. General News is represented by the daily fraction of articles containing at least three keyword occurrences (see Materials and Methods). The Improved Model (black line) leverages the past General News and Searches, together with present Searches, to infer the dynamics of General News.
}\label{fig:one}
\end{figure*}

To investigate the nature of the relation between supply and demand of news about a certain subject, six keywords, referring to the most searched subjects in Italy over the entire observation period, were selected: {\em coronavirus}, \emph{regionali}, \emph{playstation}, \emph{papa francesco}, \emph{eurovision}, \emph{sondaggi} (Supplemental Fig.~\ref{fig:descriptive})). General News and Searches for {\em coronavirus} are reported in Fig.~\ref{fig:one}. For each keyword, the time series of the daily appearances in the General News and the daily volume of queries in the Searches were simultaneously fit by Vector Auto-Regression (VAR) linear modelling ~\cite{hamilton1994timeseries}. VAR models with different lag parameters, which encapsulate the system's memory, were considered, and the best parameters were identified via the Akaike criterion~\cite{akaike1974aic} (see Materials and Methods). For all keywords, best-fitting lags ranged between 2 and 4, suggesting a typical, short-memory timescale in the system (see Supplemental Fig.~\ref{fig:causality})).

Within the VAR framework, we performed the test for Granger-causality~\cite{hamilton1994timeseries} to illustrate which time series, between General News and Searches, contributed more to the prediction of the other, and if any contribution was significant. For the majority of keywords, the contribution of past Searches to present General News was most significant (i.e. \emph{coronavirus}, \emph{regionali}, \emph{playstation}, \emph{papa francesco}) (see Supplemental Fig.~\ref{fig:causality})). We could safely assume that Searches anticipates General News and use this assumption to improve the model of the temporal behaviour of the latter. We modified the VAR equation for the evolution of General News by inserting Searches' role. More precisely, let $S(t)$ and $N(t)$ be, respectively, the values of Searches and General News at day $t$, then the new equation for the evolution of $N(t)$ reads:
\begin{equation}
 N(t) = \sum_{i=1}^{d} (\alpha_i N(t-i) + \beta_i S(t-i)) + \beta_0 S(t).
 \label{eq:VAR}
\end{equation}
where the coefficients $\alpha_i$, $\beta_0$ and $\beta_i$ were fitted, while the Akaike criterion provides the optimal lag $d$. This Improved Model closely reproduced the data, particularly in correspondence with the peaks (Fig.~\ref{fig:one} for \emph{coronavirus} and Supplemental Fig.~\ref{fig:models}). 

The model's parameters also provided a quantitative insight on the interplay between General News and Searches (Tab.~\ref{tab:modelparam}):
\begin{itemize}
 \item $\alpha_1$ was larger than other $\alpha$ parameters, indicating a strong dependence of General News on the previous day activity. This evidence is a sign of an inertial behaviour of the news supply.
 \item $\beta_0$, the weight of present Searches, was typically larger than other $\beta$ parameters and significantly non-zero, supporting the assumption of present Searches role for the Improved Model.
 \item The remaining parameters were smaller though almost always significant. For two keywords (\emph{coronavirus} and \emph{regionali}), the parameters $\beta_d$ (for $d \ge 1$) were negative. This result suggests that General News depends on the different quotient of Searches, together with the volume of Searches itself.
\end{itemize}
Of note, a direct comparison between $\alpha$ and $\beta$ parameters was not possible, as Searches and General News were scaled differently (Google Trends does not disclose the absolute scale of queries volume).

\begin{table}
\centering
\caption{The parameters and the $R^2$ resulting from the improved linear model of equation~\ref{eq:VAR} for the $4$ selected keywords. As a reference, in brackets, we report the value for $R^2$ of a trivial model with equation $N(t) = \alpha N(t-1)$, i.e., a model where every day depends only on the day before. The value of $R^2$  of the improved linear model is systematically larger than that of the trivial model. Starred values are those not significantly different from zero.}
\begin{tabular}{rcccc}
 & \emph{coronavirus} & \emph{regionali} & \emph{playstation} & \emph{papa francesco}\\
\midrule
$\alpha_1$ & $0.82$ & $0.65$ & $0.18$ & $0.54$ \\
$\alpha_2$ &  & $0.26$ & $0.19$ &  \\
$\beta_0$ & $0.070$ & $0.0082$ & $0.00055$ & $0.0038$ \\
$\beta_1$ & $-0.034$ & $0.003*$ & $0.00035$ &  \\
$\beta_2$ &  & $-0.0064$ & $0.00068$ &  \\
\midrule
$R^2$ & $0.996$ $(0.991)$ & $0.89$ $(0.86)$ & $0.54$ $(0.29)$ & $0.73$ $(0.63)$ \\
\bottomrule
\end{tabular}
\label{tab:modelparam}
\end{table}

\subsection*{The different behaviours of General News and Fake News}

The Improved Model quantifies the information supply dynamics and enables the comparison between General News and disinformation supply. We applied this methodology to the topic \emph{coronavirus}, since it dominated the landscape of information (Supplemental Fig.~\ref{fig:descriptive}), and due to the direct impact of disinformation on the response to the 2020 pandemic. To this end, we extended our analysis to the news items that were annotated as false or misleading, thus named Fake News (see Material and Methods).




We exploited the Improved Model~\ref{eq:VAR} to compare General News and Fake News through their best-fitting coefficients $\alpha$ and $\beta$. 
To this end, we paralleled the variable $N(t)$, the daily proportion of \emph{coronavirus}-related General News at day $t$, and $FN(t)$, the daily proportion of \emph{coronavirus}-related Fake News at day $t$ (Tab.~\ref{tab:compa}). 

Compared to General News, \emph{coronavirus}-related Fake News shows a meaningfully lower Inertia term, $\alpha_1$, and a non-significant $\beta_1$ indicating a greater reactivity to $S(t)$. These pieces of evidence and the lower prediction score (adjusted $R^2$) suggest that disinformation presents a different behaviour than General News, to the points that it distorts the dynamics of the news ecosystem and leads to impaired modelling performance.  

\begin{table}[ht]
\centering
\caption{Coefficients from the Improved Model fitting of General News and of Fake News having at least one occurrence of the keyword {\em coronavirus} (see Materials and Methods). Starred coefficients do not differ significantly from zero.}
\begin{tabular}{lrr}
 & General News & Fake News \\
\midrule
$\alpha_1$ (Inertia) & $0.860 \pm 0.016$ & $0.758 \pm 0.039$ \\
$\beta_0$ & $0.460 \pm 0.035$ & $0.294 \pm 0.086$ \\
$\beta_1$ & $-0.248 \pm 0.042$ & $-0.081^{*} \pm 0.091$ \\
\midrule
$R^2$ & $0.995$ & $0.931$ \\
\bottomrule
\end{tabular}
\label{tab:compa}
\end{table}

Another difference in the behaviour of news and disinformation emerged at a semantic level. We focused on the most queried keywords searched together with \emph{coronavirus} in Google Search (see Materials and Methods). Each of these related queries provided a time series of news demand about a sub-domain that co-occurs with, and therefore is semantically linked to, \emph{coronavirus}. We quantified the co-occurrence of these terms with the \emph{coronavirus} keyword also in the news items, for both General News and Fake News. In this way, we defined $\mathbf{S}(t), \mathbf{N}(t), \mathbf{FN}(t)$ as the daily semantic vectors for \emph{coronavirus}-related Searches, General News, and Fake News, respectively. Each vector has seventeen entries, one per sub-domain (see Materials and Methods for more details).

\begin{figure*}[ht]
\centering
\includegraphics[width=.9\textwidth]{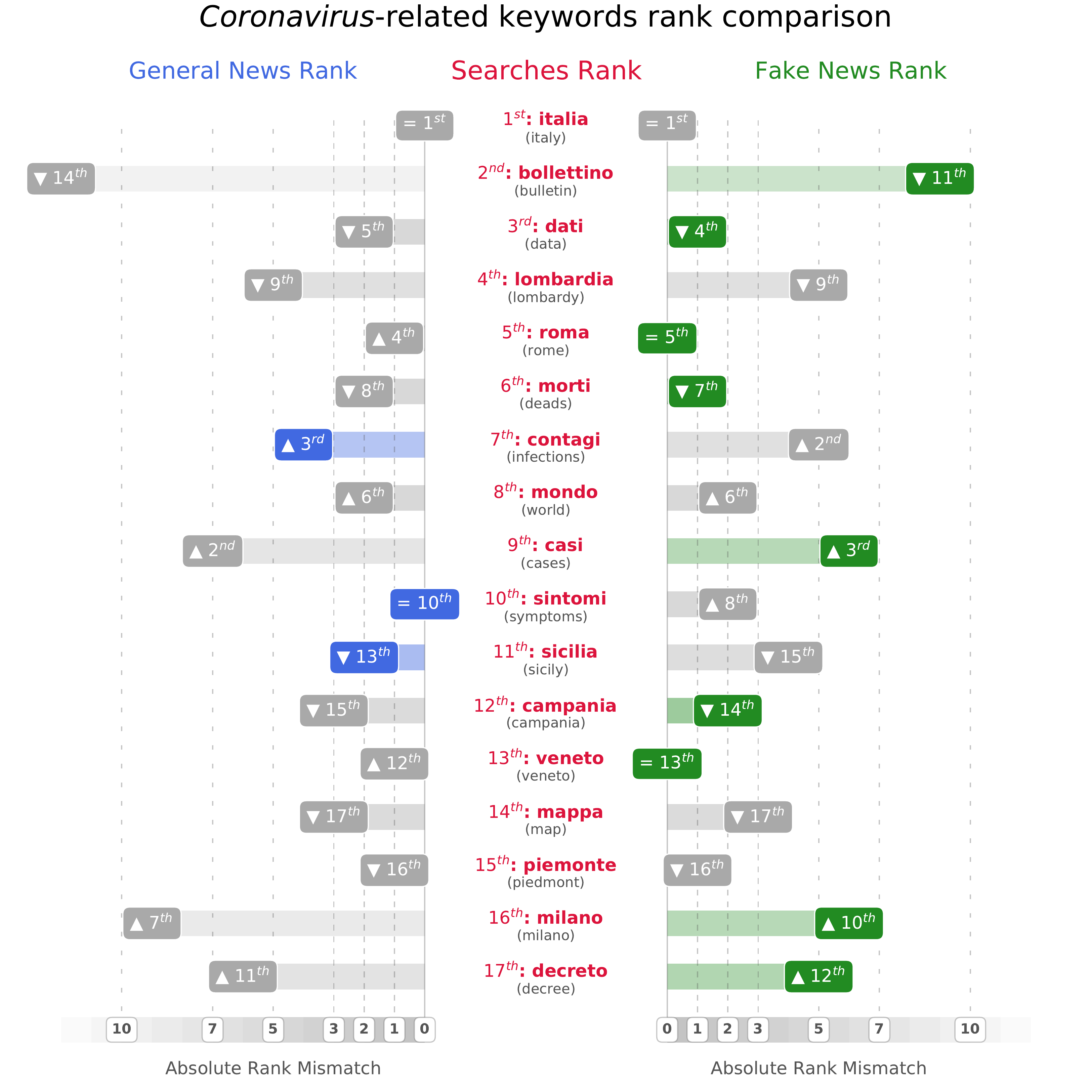}
\caption{The ranked components of $\mathbf{S}_{tot}$ (centre red), representing \emph{coronavirus} sub-domains sorted by total news demand over the observed time. 
On the sides of each keyword, a tag indicates the rank in $\mathbf{N}_{tot}$ for General News, on the left, and in $\mathbf{FN}_{tot}$ for Fake News, on the right.
Tags are distanced from the centre by the amount of rank mismatch to Searches ranks. Tags are coloured to highlight the rank closest to the Searches rank: blue for General News and green for Fake News. 
}\label{fig:two}
\end{figure*}

We calculated $\mathbf{S}_{tot}=\sum_t\mathbf{S}(t)$ and sort its components to rank the different sub-domains by the total news demand over the period considered (Fig.~\ref{fig:two}).
To assess the difference between information and disinformation with respect to the matching of news demand for different sub-domains, we challenged the components' rankings of  $\mathbf{N}_{tot}=\sum_t\mathbf{N}(t)$ and $\mathbf{FN}_{tot}=\sum_t\mathbf{FN}(t)$ against the corresponding ones of $\mathbf{S}_{tot}$ (Fig.~\ref{fig:two}).


Given the \emph{coronavirus}-related keywords ranked from the Searches as a reference, Fake News ranking shows fewer and minor mismatches compared to General News. We quantified this difference in behaviour through Spearman's Correlation. $\mathbf{S}_{tot}$ and $\mathbf{N}_{tot}$ components resulted positively correlated ($r = 0.52$, with a p-value of $0.031$) but $\mathbf{S}_{tot}$ and $\mathbf{FN}_{tot}$ correlated more ($r = 0.67$, with a p-value of $0.0032$).

The semantic difference in the behaviour of Fake News and General News holds not only at the aggregated level but also at a daily level. This was measured through the cosine distance $\textrm{d}(\cdot, \cdot)$ on their daily vectors $\mathbf{S}(t)$, $\mathbf{N}(t)$ and $\mathbf{FN}(t)$ (see Materials and Methods). Again, Searches were taken as reference and we calculated its cosine distance from General News, $\textrm{d}(\mathbf{S}(t), \mathbf{N}(t))$, and from Fake News, $\textrm{d}(\mathbf{S}(t), \mathbf{FN}(t))$. The daily relative difference  between the cosine distances of Searches-Fake News and Searches-General News
\begin{equation}
\frac{\textrm{d}(\mathbf{S}(t), \mathbf{FN}(t))-\textrm{d}(\mathbf{S}(t), \mathbf{N}(t))}{\textrm{d}(\mathbf{S}(t), \mathbf{N}(t))}
\end{equation}
resulted in negatives values in most days $t$ (Supplemental Fig.~\ref{fig:cosine})). In fact, both the Mean ($-0.13$) and Median ($-0.15$) were negative, indicating that the cosine distance Searches-Fake News is generally smaller than that of Searches-General News. This result shows how Fake News meets news demand better than General News. 

\subsection*{Independent detection of Fake News concentration}

The observed differences between General News and Fake News dynamics can be exploited to assess disinformation about the topic {\em coronavirus}.

The difference in modelling Fake and General News suggests that when Fake News concentration on a topic rises, the General News dynamics, which includes Fake News, becomes perturbed. We hypothesise that this perturbation is expected to impair the General News modelling performance. To test this hypothesis, the Improved Model was fit to General News locally on a time window of 14 days, sliding over the entire data time range (see Materials and Methods). For each window, centred in $t$, we computed the local modelling error defined as:
\begin{equation}
E(t) = (1-R^2(t))\cdot \langle N\rangle (t), 
\end{equation}
where $R^2(t)$, the $R^2$ score for the model fitted to the window, is weighted by $\langle N\rangle (t)$, the average volume of news produced in that time window.

Although formulated without exploiting disinformation annotations, $E(t)$ significantly correlates with the concentration of Fake News on the \emph{coronavirus} subject, $FN(t) / N(t)$ (Spearman's $r=0.47$, with a p-value of $3.9\cdot 10^{-13}$ (see Materials and Methods). This result supports the hypothesis that loss of predictability from the General News dynamics co-occurs with disinformation spikes. As a consequence, $E(t)$ stands as a very promising proxy for the concentration of disinformation about the topic \emph{coronavirus}.

The semantic difference between General News and Fake News suggests that disinformation suppliers might react not only to the news demand but, in particular, to the {\em semantically unsatisfied} news demand. We hypothesized that as General News becomes more semantically distant from Searches, Fake News would fill that gap. This hypothesis was tested all over the time range measuring the daily cosine distance between the semantic vectors of Searches and General News $K(t) = \textrm{d}(\mathbf{S}(t),\mathbf{N}(t))$. We then checked the correlation of $K(t)$ with the daily concentration of Fake News, $FN(t)/N(t)$, about {\em coronavirus}. The correlation turns out to be positive and significant (Spearman's $r=0.58$, with a p-value of $1.6\cdot 10^{-21}$), supporting the hypothesis. This result allows us to adopt $K$ as a second independent indicator for Fake News concentration assessment. 

\begin{figure*}[ht]
\centering
\includegraphics[width=.8\textwidth]{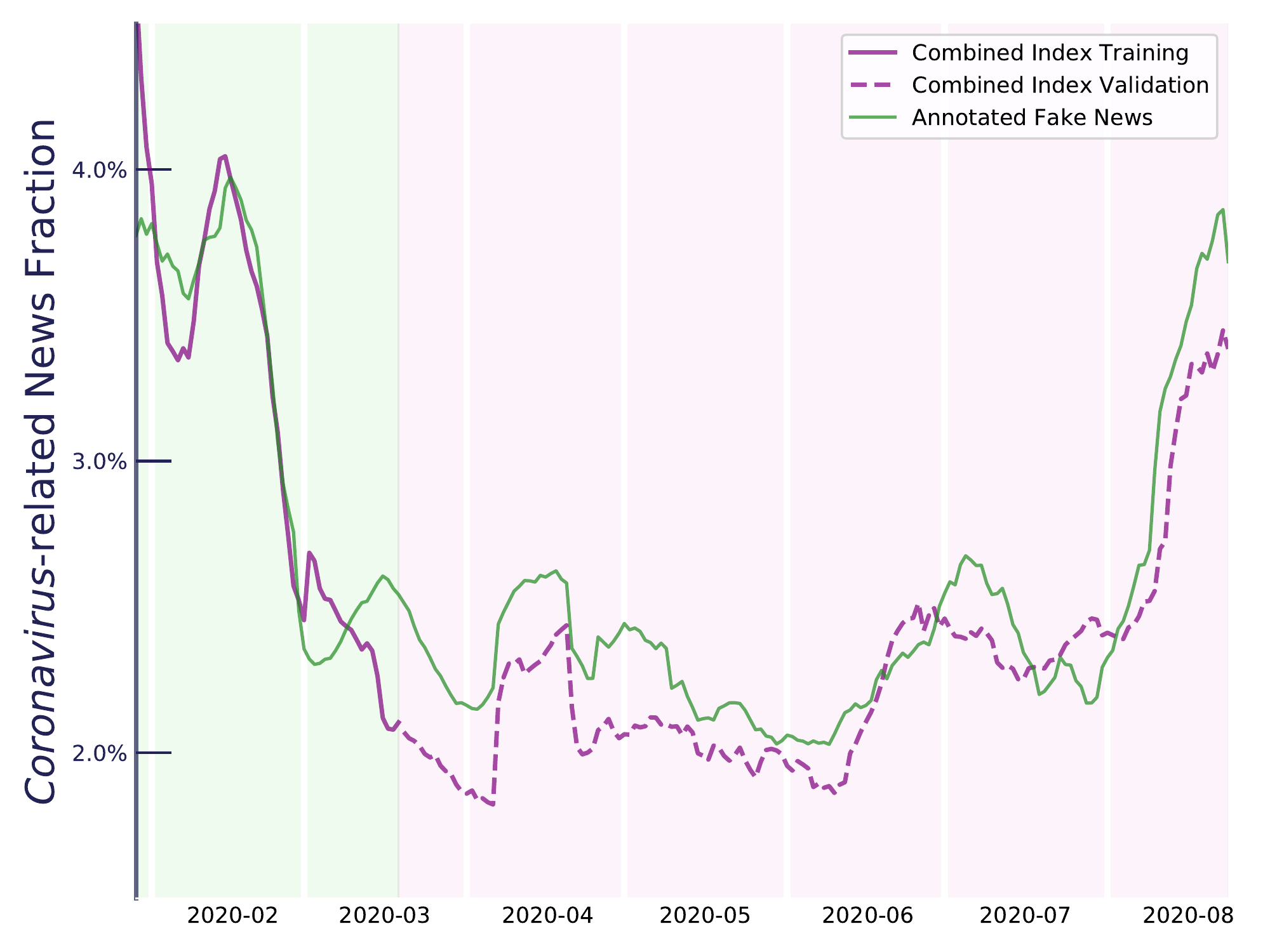}
\caption{The time series of news annotated as Fake, normalised through the total number of  \emph{coronavirus}-related News compared with the Combined Index for disinformation. The Combined Index is defined as a linear combination of the weighted modelling error for the local fitting of News within the improved Vector Auto-Regression model and the cosine distance between the semantic vectors of Searches and News. The parameters of the combination were fitted in the training set and then tested in the validation set.}\label{fig:three}
\end{figure*}

To test the effectiveness of our indicators $E$ and $K$ in assessing disinformation concentration, we combined them in a Combined Index for disinformation (see Material and Methods). We fit them linearly on a training set composed of approximately the first $25\%$ of data from the time series, providing the best linear combination of the two (see Fig.~\ref{fig:three}).

The Combined Index was then tested against the validation set, achieving substantial accuracy (reduced chi-squared statistic of $0.945$). All these findings suggest that the Combined Index provides a valuable measure for detecting disinformation concerning SARS-CoV-2.

The methodology could also be applied to different topics, and to the aggregation of several topics, to assess the health status of the news ecosystem at a more general level. To challenge this claim, we considered the set of the $4$ keywords modelled. We aggregated them to create a synthetic macro-topic, for which they individually represented the analogous of the related queries we have seen before. We judged the adoption of the first indicator, i.e. the weighted modelling error for the local fitting, to be pointless since the macro-subject dynamics is largely dominated by the topic \emph{coronavirus}. This would have resulted in an indicator similar to the modelling of the \emph{coronavirus} component alone. We thus focused only on the second indicator, i.e., the cosine distance between the semantic vectors of Searches and News, where the components of the vectors are now the values of General News, Fake News, and Searches for the $4$ keywords. The daily value of cosine distance between General News and Searches of the synthetic subject correlates positively and meaningfully with the concentration of disinformation on the synthetic subject (Spearman's correlation of $0.44$ with p-value $ = 1.8\cdot 10^{-11}$). This result supports the plausibility of the application of our methodology in wider contexts.

\section*{Discussion}

Information quality is a fundamental challenge for the Information Age, especially during a pandemic. Studying the general news system and comparing  it with the subset of news labelled as disinformation, we found that pandemic-related Fake News production seems more reactive and precise than General News supply in addressing people's news demand. We exploited such a difference to develop an index for vulnerability of specific topics to disinformation takeover.

The analysis of Searches and General News for \emph{coronavirus} and a set of other \emph{coronavirus}-unrelated highly queried keywords exposed the relation between supply and demand of news: (i) a linear modelling scheme was effective in almost all cases; (ii) the memory of the process seems to be very short (2-4 days) in all cases; (iii) causality was more commonly directed from Searches to General News (e.g. for \emph{coronavirus}). Thanks to these considerations, an improved descriptive model could be developed to better describe the relationship between supply and demand for information. This modelling framework allowed us to discern how the inertia of news suppliers is one of the main traits of the dynamics for all the studied keywords. Also, the negative dependence on previous days Searches observed in some cases suggests a dynamics where the trend of the interest is more important than the interest itself for news producers.

The comparison of \emph{coronavirus}-related General News and Fake News through the improved linear model's lens exposed that Fake News feature lower inertia and a different dependence on Searches, quantifying their more reactive behaviour. We can speculate that this behavioural difference could be a consequence of the different production environments of General News and Fake News. The firsts are mainly produced by a large and well-established community of professional journalists while the latter are the outcome of by a scattered multitude of small, unorganized actors. The community size effect might be responsible for the different inertial behaviour observed.

The semantic analysis revealed another key difference between the dynamics of General News and Fake News. Looking at the shares of the most queried keywords co-occurring with \emph{coronavirus} we discovered that Fake News is better aligned to Searches than General News not only at a cumulative level but also daily over the entire observation period. This result suggests that disinformation tends to be more semantically precise than information in matching people's interests. This difference might be explained considering the different aims of the two communities. While they are both interested in answering people's demand for information, general news producers also have the ambition for complete coverage of topics, while Fake News producers can focus on chasing the people's attention.

We exploited the modelling and the semantic mismatch between General News and Fake News to introduce two indexes to detect bursts of disinformation on \emph{coronavirus}. It is worth mentioning how these indexes do not rely on, in their definition, any information about Fake News. They are based instead only on the time series of General News and Searches. The first index is based on the modelling of General News and Fake News and exploits the goodness of the modelling scheme. Since General News includes Fake News, a higher presence of the latter could be revealed by a worse performance of the modelling scheme, quantified through the local weighted modelling error. A higher value of this indicator means that the normal relations between General News and Searches have been altered, presumably by the presence of Fake News. The second index exploits the semantic daily misalignment between General News and Searches. In this case, a higher value of this indicator signals that semantic imprecision of General News leaves readers' interests unsatisfied, possibly fostering the success of Fake News.

The positive and meaningful correlation of both indicators with Fake News concentration on \emph{Coronavirus} supports two hypotheses. The first is that disinformation perturbs the normal interplay between General News and Searches. The second one is that disinformation is fuelled by the semantic misalignment between General News and Searches.

The two indexes discussed above blend into a single Combined Index for disinformation. We adopted a training set for its definition and a validation set to test its performances. Thanks to its independence of Fake News annotations and its potential generalisation to other topics, the Combined Index can be a powerful tool for journalists and editors, on the one hand, and news monitoring authorities, on the other, to detect in real-time vulnerabilities to disinformation. Our results also suggest, as a possible strategy to face these vulnerabilities, a timely refocus of General News supply to better meet the information demand of the public.

Information vulnerabilities are a major risk factor for our societies, as they directly impact individuals in their behaviours and choices.  For example, the solution to the Coronavirus crisis heavily depends on individuals' behaviours, which, in turn, are directly affected by the news. The approach presented here, far from being conclusive, represents a first contribution towards a deeper understanding of the phenomenology of disinformation as part of the information ecosystem's general dynamics. Additional studies will be needed to test the conclusions and to generalise the results to different countries, languages, domains, and time periods. Moreover, the diffusion layer should be added to the analysis of the dynamics of the infosphere, with particular attention to the social media spreading of news. In our opinion, a paradigm shift in facing disinformation is no more an option. Instead, it is a pressing need, and we contend that the work we presented may contribute to the shift of scientific research towards a more concrete view, aiming to provide policymakers with knowledge and tools to prevent and fight disinformation. 


\section*{Materials and Methods}
\subsection*{Searches Data}

The information demand about a specific subject was obtained from Google Trends, a platform providing access to an anonymous sample of actual search requests made in Google Search engine, from a selected location and time interval.

For each given keyword, Google Search returns a time-series with values proportional to the number of times the keyword was searched each day. Since Google Search does not disclose the actual number of searches, the time series values are rendered as percentages of the maximum number returned. As a result, data consist of integers ranging in the interval $(0, 100)$.  The time series of one keyword was referred to as the ``Searches'' for that keyword and provided a measure of the interest it received.

The use of ``pytrends'' library for Python\footnote{\url{https://github.com/GeneralMills/pytrends}} enabled the interaction with the Google Trends platform. The terms from Supplemental Fig.~\ref{fig:descriptive} were requested separately, for the time ranging from the 6th of December 2019 to the 31st of August 2020 in Italy. These were: \emph{coronavirus}, \emph{regionali} (regional elections), \emph{playstation}, \emph{papa francesco} (Pope Francesco), \emph{eurovision} (the European music contest), \emph{sondaggi} (polls).

Google Trends also provided information about queries most searched with a specific keyword. In particular, the most popular queries related to the keyword \emph{coronavirus} (e.g.,  \emph{ coronavirus news} ) were gathered. Such list is capped by Google Trends at a maximum of 25 related keywords, ordered by most searched to least, and denoted $q_1(t), \ldots, q_{25}(t)$ respectively ($t$ indicating the time), with $q_0(t)$ the time series of \emph{coronavirus} searches.

To compare the searches of a given keyword with its related keywords, it is necessary to put them on the same scale. To this end, searched items were queried in pairs. In this way, Google Trends normalized the two resulting time series for the highest of the maximums of the two. Given the two times-series per request $(q_{i-1}(t), q_{i}(t))$, with $i=1,\ldots,25$, a coefficient $\alpha_{i} = max_t (q_{i-1}(t)) / max_t (q_{i}(t))$ was calculated. Thus, all the time series $q_i$ could be set on the same scale of $q_0$, multiplying by $\prod_{j=1}^i \alpha_j$. This procedure is needed not to lose resolution on keywords with a small number of queries. Having queried for pairs $(q_{0}(t), q_{i}(t))$, would have resulted in a rounding at $0$ performed by Google Trends.

{\em coronavirus}-related queries were then aggregated by summing up their time-series. Thus, \emph{coronavirus oggi} (Coronavirus today), \emph{coronavirus notizie} (Coronavirus news), \emph{coronavirus ultime} (Coronavirus latest), \emph{coronavirus ultime notizie} (Coronavirus latest news) and \emph{coronavirus news}, were all aggregated into \emph{coronavirus news}. Subsequently, we removed all the queries that returned the same search results as another query. These were \emph{coronavirus contagi} (Coronavirus infections) and \emph{coronavirus in italia} (Coronavirus in Italy), duplicates of \emph{contagi coronavirus} (Coronavirus infections), and \emph{coronavirus italia} (Coronavirus Italy), respectively. Also, the query \emph{corona} was excluded because it has other meanings in Italian, namely ``crown'', and it is also a famous brand of beer.
Finally, the list of queries associated with \emph{coronavirus}, ordered by the amount of searches, was: \emph{news}, \emph{italia} (Italy), \emph{lombardia} (Lombardy), \emph{sintomi} (symptoms), \emph{contagi} (infections), \emph{casi} (cases), \emph{morti} (deaths), \emph{bollettino} (bulletin), \emph{roma} (Rome), \emph{dati} (data), \emph{mondo} (world), \emph{mappa} (map), \emph{sicilia} (Sicily), \emph{veneto}, \emph{campania}, \emph{decreto} (decree), \emph{milano} (Milan), \emph{piemonte} (Piedmont).

\subsection*{News Data}

To analyse the news supply, we investigated the data provided by AGCOM, the Italian Authority for Communications Guarantees, which granted us access to the content of a vast number of Italian news sources published online and offline from the 6th of December 2019 to the 31st of August 2020 in Italy. These data included articles from printed and digital newspapers and information agencies, TV, radio sites, and scientific sources. Moreover, the data had a specific annotation for ``fake news'' sources. The list of disinformation sources, vetted by the Authority,  is provided by fact-checking organizations like {\href{https://www.bufale.net/}{bufale.net}} and {\href{https://www.butac.it/}{butac.it}}. It had already been employed for other studies on disinformation~\cite{vicario2019polarization}.

After pre-processing the data for duplicates and incomplete logs elimination, the General News data consisted of almost 7 million entries from 554 different news sources. Each data entry has a unique ID and contains, among other information, the title and the content of the piece of news, its date, its source, and the annotation of belonging to the disinformation sources list). 

Needing to imitate the rationale underlying Google Trends data, where daily search counts refer to the query of specific keywords, we sought to find counts of daily keywords also in the news data. To do so, given a keyword (e.g., \emph{coronavirus}), we defined three different metrics: the piece of news containing the keyword at least once, those having the keyword at least three times, and finally, all the occurrences of a specific keyword. These three metrics were then normalized on the total number of news' sources per day to level the press activity during weekends. For each model, we chose the metric with the best modelling performances. For the improved version of the VAR model described in equation~\ref{eq:VAR} from the Results section, the metric with at least three occurrences was selected, even if the other two showed similar performances. Instead, the most inclusive metric (at least one occurrence) was adopted when dealing with disinformation. This procedure was necessary to enhance the signal, given the low number of Fake News items encountered. For consistency, the overall General News was considered with the same metric (at least one occurrence) when comparing it with the Fake News time series.

Following the same rationale, we adopted the first metric to filter for the keywords related to the \emph{Coronavirus} subject described in the previous subsection. To do so, we selected the piece of news containing the keyword \emph{coronavirus} at least once, and, in this subset, we counted the ones featuring the desired related keyword at least once. The values found were normalised on the total number of news pieces having the keyword \emph{coronavirus} at least once per day to get a proxy for the share of \emph{Coronavirus} piece of information focused on the related keyword sub-domain. We repeated this analysis for the subset of news mentioning the keyword \emph{coronavirus} at least once and marked as Fake News in the data. We then used the values extracted from this analysis to investigate the disinformation supply in the \emph{Coronavirus} context.

\subsection*{Time Series Analysis}
Time series of Searches and General News, from Supplemental Fig.~\ref{fig:descriptive}, were investigated using the VAR model~\cite{hamilton1994timeseries}, using Python's {\em statsmodels} package for time series analysis \cite{seabold2010statsmodels}. Data were regularized via $x \mapsto log(1+x)$ transformation before fitting. For the VAR modelling, the number of lags $d$ was determined as the parameter that minimized the Akaike information criterion \cite{akaike1974aic}, with $d$ ranging in the interval $(1,14)$. This modelling strategy was chosen to ensure the interpretability of the fitted model and its regression coefficients.

From the VAR model, we computed Granger-causality~\cite{hamilton1994timeseries} to test whether queries' values provided meaningful information to the prediction of news volumes and vice versa. Since two tests were performed on the same data from a given subject (for the null hypotheses ``Searches do not Granger-cause General News'' and ``General News do not Granger-cause Searches''), resulting p-values were corrected by the Holm-Bonferroni method~\cite{lehmann2006testing} Thus, pairs of p-values in Supplemental Fig.~\ref{fig:causality} were multiplied by $2$, to control for family-wise error rate and to maintain comparability.

In Fig.~\ref{fig:one} and Supplemental Fig.~\ref{fig:models}, the Improved Models for regression of the General News were derived adjusting the VAR models to include Searches at time $t$ (Supplemental Fig.~\ref{fig:models}). Lags were re-elaborated through the Akaike criterion as before, with similar results. These models were then compared against a null model that forecasts one day proportionally to the value of the day before to benchmark how beneficial the addition of regressing variables was to General News's prediction (see Tab.~\ref{tab:modelparam}).

To assess the semantic misalignment between General News and Searches from Supplemental Fig.~\ref{fig:cosine},
the cosine distance was calculated as $\textrm{d}(\mathbf{S}(t), \mathbf{N}(t)) = 1 - \mathbf{S}(t) \cdot \mathbf{N}(t) / \vert \mathbf{S}(t) \vert \, \vert \mathbf{N}(t) \vert$, on the vectors $\mathbf{S}(t) = (S_1(t), \ldots, S_k(t))$, $\mathbf{N}(t) = (N_1(t), \ldots, N_k(t))$ where $S_i(t)$ and $N_i(t)$ represented the volumes of searches and news, respectively, at time $t$ for the $i$-th keyword associated to {\em coronavirus}, with $\cdot$ being the dot product and $\vert \cdot \vert$ the Euclidean norm. Cosine distance was suitable to compare high-dimensional vectors at different scales, and returned values in $(0,1)$ for vectors with non-negative entries such as $\mathbf{S}(t)$ and $\mathbf{N}(t)$.

\subsection*{Combined Index Validation}
To define and validate the Combined Index from Fig.~\ref{fig:three}, we split the daily data from Fake News concentration on \emph{coronavirus} into a training set (from the 29th of January 2020 to the 20th of March 2020) and a validation set (from the 21st of March 2020 on). 

Thus, we defined the Combined Index as a linear combination of the two starting indexes that best fitted the Fake News concentration, using a linear model with Gaussian noise on the training data. The ordinary least squares estimate $\hat{\sigma}$ for the variance of the Gaussian noise was then calculated as the Mean Squared Error (MSE) divided by the statistical degrees of freedom $k$ (i.e., the number of observations minus $2$, the number of parameters in the model).

To assess the predictive potential of the Combined Index, we adopted the trained model to forecast the concentration of Fake News in the validation set. This prediction's goodness was tested through the reduced chi-squared statistic, which is calculated as the MSE on the validation set divided by $\hat{\sigma}$. This statistic is approximately distributed as a $\chi^2$ with as many degrees of freedom as the size of the validation set (i.e., 51), leading to a p-value of about $0.945$. As such, the null hypothesis, that the concentration of Fake News for the keyword \emph{coronavirus} is distributed in agreement with the trained model, cannot be rejected.



\section*{Acknowledgements}
The Authors wish to warmly thank Marco Delmastro of AGCOM for insightful discussions about the Italian news ecosystem as well as for providing the database of Italian news. The database was shared in the framework of the Task Force on "Digital Platforms and Big Data - Covid-19 Emergency", established by the AGCOM to contribute, among other things, to the fight against online disinformation on issues related to the SARS-CoV-2 crisis.


\bibliographystyle{plain}

\bibliography{biblio}

\pagebreak
\begin{center}
\textbf{\Large { \mytitle} }
\end{center}
\begin{center}
\textbf{\LARGE Supplementary Information}
\end{center}
\setcounter{equation}{0}
\setcounter{figure}{0}
\setcounter{table}{0}
\setcounter{page}{1}
\makeatletter
\renewcommand{\theequation}{S\arabic{equation}}
\renewcommand{\thefigure}{S\arabic{figure}}
\renewcommand{\theHtable}{Supplement.\thetable}
\renewcommand{\theHfigure}{Supplement.\thefigure}


\begin{figure*}[ht]
\centering
\includegraphics[width=1.\textwidth]{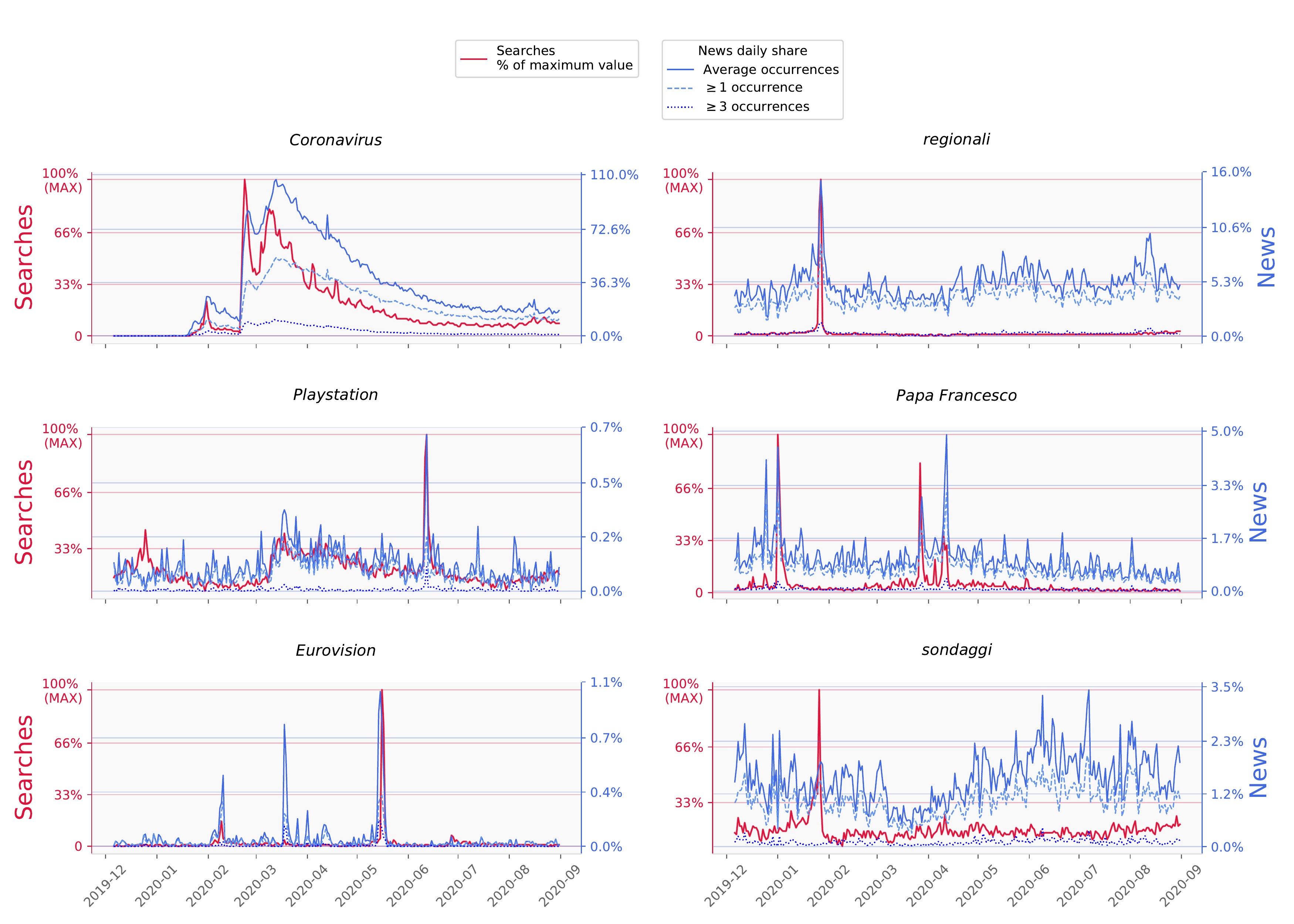}
\caption{Temporal behaviour of the fractions of Searches (red, left $y$-axis) and General News (blue, right $y$-axis), in Italy from early December 2019 to the end of August 2020, about six selected keywords. Searches are reported as a percentage of the maximum observed in the monitored period. General News are represented instead by three curves showing: the daily fraction of articles containing at least one occurrence of the keyword (dotted line), the daily fraction of articles containing at least three occurrences of the keyword (dashed line), and the average number of occurrences per article (continuous line). The six keywords are \emph{coronavirus}, \emph{playstation}, \emph{ eurovision}, \emph{regionali} (regional elections), \emph{ papa francesco} (Pope Francesco) and  \emph{sondaggi} (polls). \emph{coronavirus} was dominant in the information landscape of the observed period, resulting to be the most queried term and and the most frequent keyword in the news. The peak in mid-march, in particular, shows that, on average, every newspaper article mentioned it once.
}\label{fig:descriptive}
\end{figure*}

\begin{figure}
\centering
\includegraphics[width=.7\textwidth]{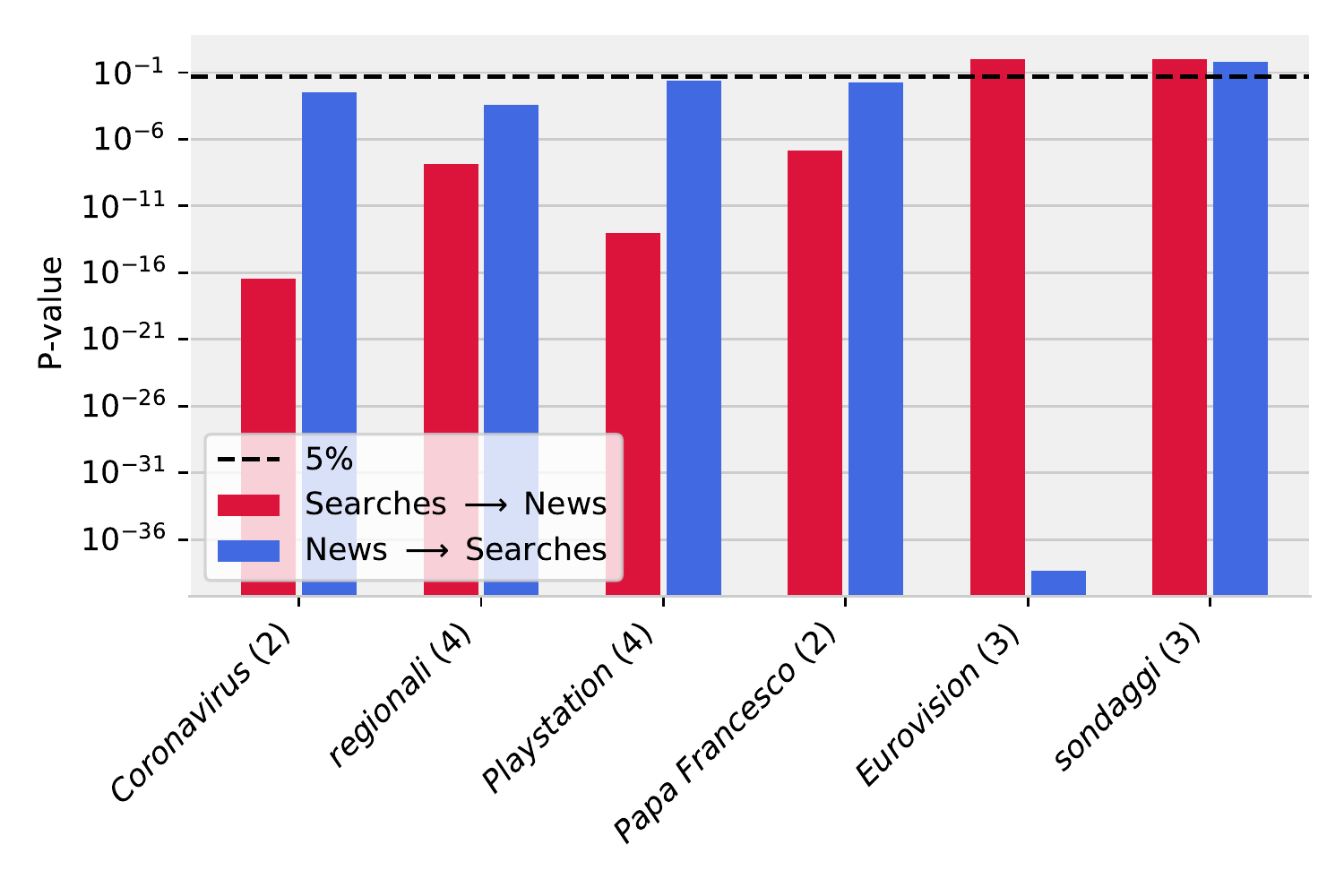}
\caption{The results of the Granger-causality test for the six keywords investigated. We report the $p$-values and, next to each keyword, the time windows' size for the best model in brackets. The horizontal black line highlights the 5\% significance value. When the bars go above such a threshold, the corresponding test for Granger-causality is not significant. Lower bars suggest a stronger causality. The majority of the most queried keywords shows a Search Granger-caused dynamics. }
\label{fig:causality}
\end{figure}

\begin{figure}
\centering
\includegraphics[width=.6\textwidth]{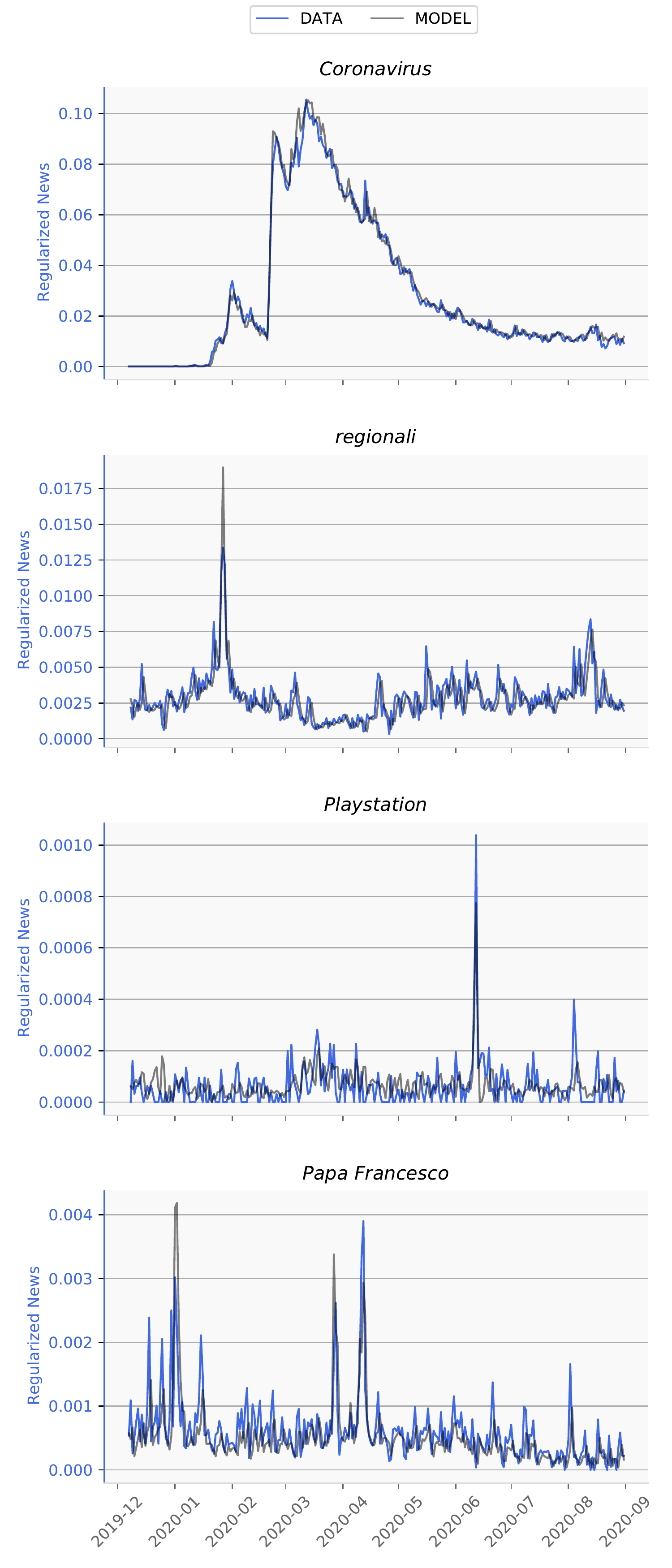}
\caption{Comparison of the time series for General News and the outcome of the Improved Model from equation~\ref{eq:VAR}. Numeric results are reported in Tab.~\ref{tab:modelparam}}
\label{fig:models}
\end{figure}

\begin{figure}
\centering
\includegraphics[width=.7\textwidth]{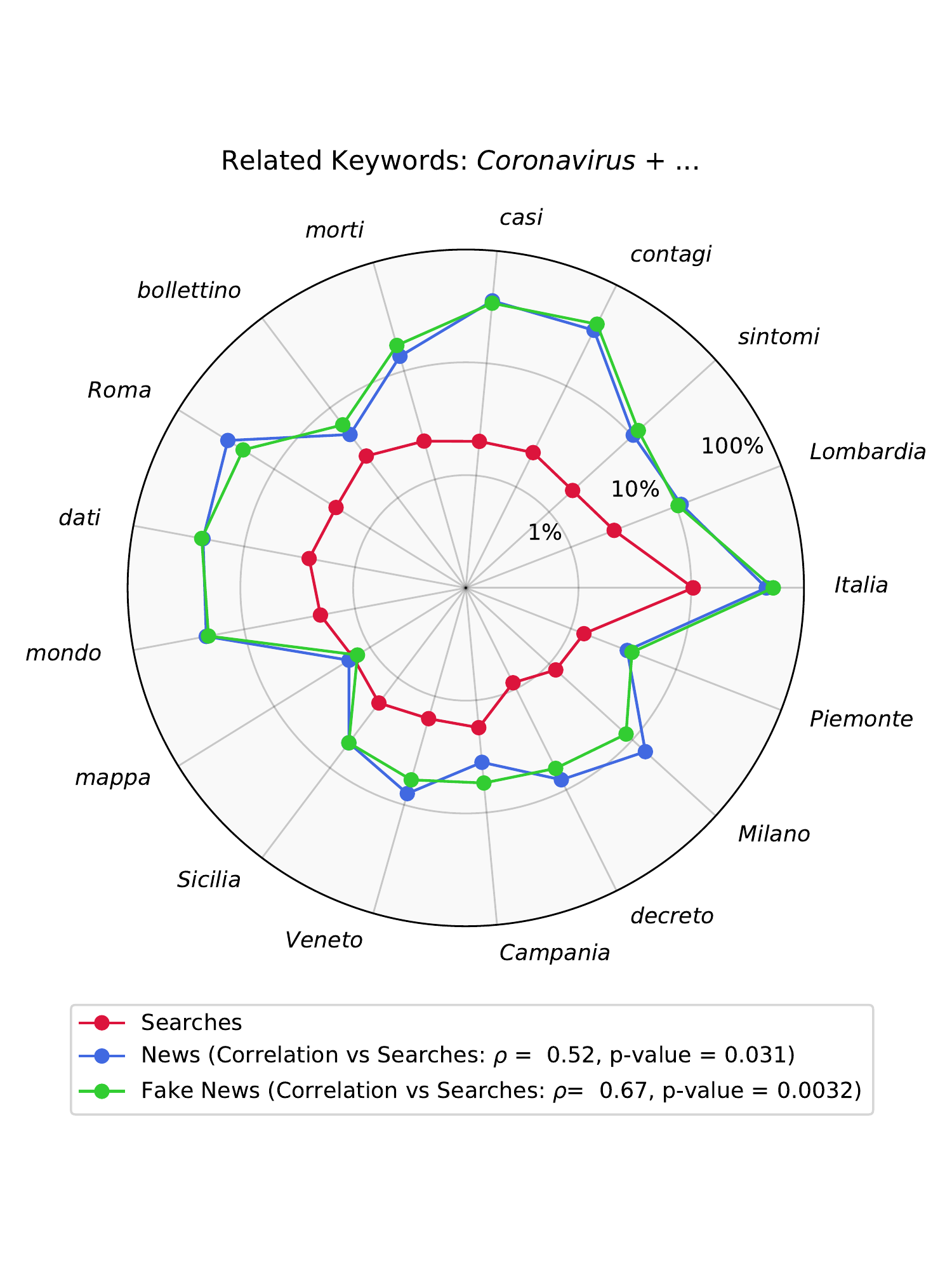}
\caption{Radar representation of the total semantic vectors for Searches, General News, and Fake News. The radius is the logarithm of the average percentage of the related keyword. The results from Spearman's correlations are described in the legend. Searches are normalized to $1$, while General News and Fake News sum to more than $1$ given the metrics' non-exclusivity (a piece of news can be counted in more sub-domains if it contains more keywords). For the analysis performed, however, the scales are irrelevant.}
\label{fig:radar}
\end{figure}
\begin{figure}
\centering
\includegraphics[width=.7\textwidth]{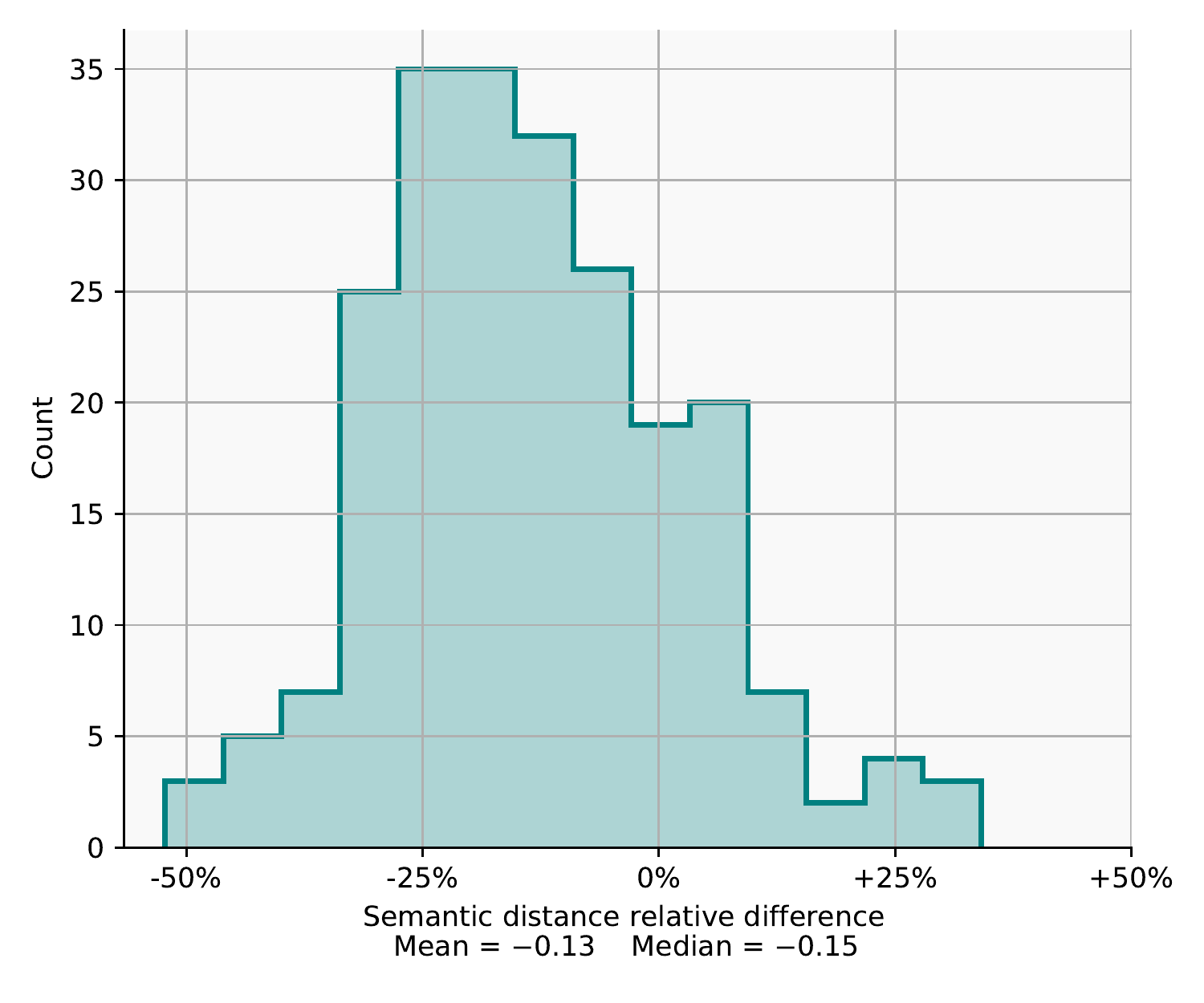}
\caption{The histogram of the daily relative differences between the cosine distance Searches-Fake News and the cosine distance Searches-General News. Both average and median are below zero, indicating that Fake News are closer to Searches than General News even on daily basis.}
\label{fig:cosine}
\end{figure}

\end{document}